\documentclass[reprint,aps,prl,superscriptaddress,]{revtex4-2}
\usepackage[utf8]{inputenc}
\usepackage{graphicx}
\usepackage{amssymb}
\usepackage{amsmath}
\usepackage{bm}
\usepackage[euler]{textgreek}

\usepackage{jabbrv}

\newcommand{\dt}{\Delta t}

\newcommand{\prob}[1]{\mathcal{P}\left(#1\right)}

\newcommand{\normal}[1]{\textbf{N}\left(#1\right)}

\newcommand{\mtx}[1]{\bm{#1}}

\usepackage{color}
	
\definecolor{orange}{rgb}{1,0.5,0}

\usepackage[colorlinks=true,linkcolor=blue,citecolor=red]{hyperref}

\begin{document}
\preprint{AIP/123-QED}

\title[Inferring potential landscapes from noisy trajectories]{Inferring potential landscapes from noisy trajectories}

\author{J.~Shepard Bryan IV*}
\affiliation{Department of Physics, Arizona State University, USA}

\author{Prithviraj Basak*}
\affiliation{Department of Physics, Simon Fraser University, CA}

\author{John Bechhoefer}
\email{johnb@sfu.ca}
\affiliation{Department of Physics, Simon Fraser University, CA}

\author{Steve Press{\'e}}
\email{spresse@asu.edu}
\affiliation{Department of Physics, Arizona State University, USA}
\affiliation{School of Molecular Sciences, Arizona State University, USA}

\date{\today}

\begin{abstract}
While particle trajectories encode information on their governing potentials, potentials can be challenging to robustly extract from trajectories. Measurement errors may corrupt a particle's position, and sparse sampling of the potential limits data in higher-energy regions such as barriers. We develop a Bayesian method to infer potentials of arbitrary shape alongside measurement noise. As an alternative to Gaussian process priors over potentials, we introduce structured kernel interpolation to the Natural Sciences which allows us to extend our analysis to large data sets. Our method is validated on 1D and 2D experimental trajectories for particles in a feedback trap.
\end{abstract}

\maketitle

\begin{figure*}
    \centering
    \includegraphics[width=\textwidth]{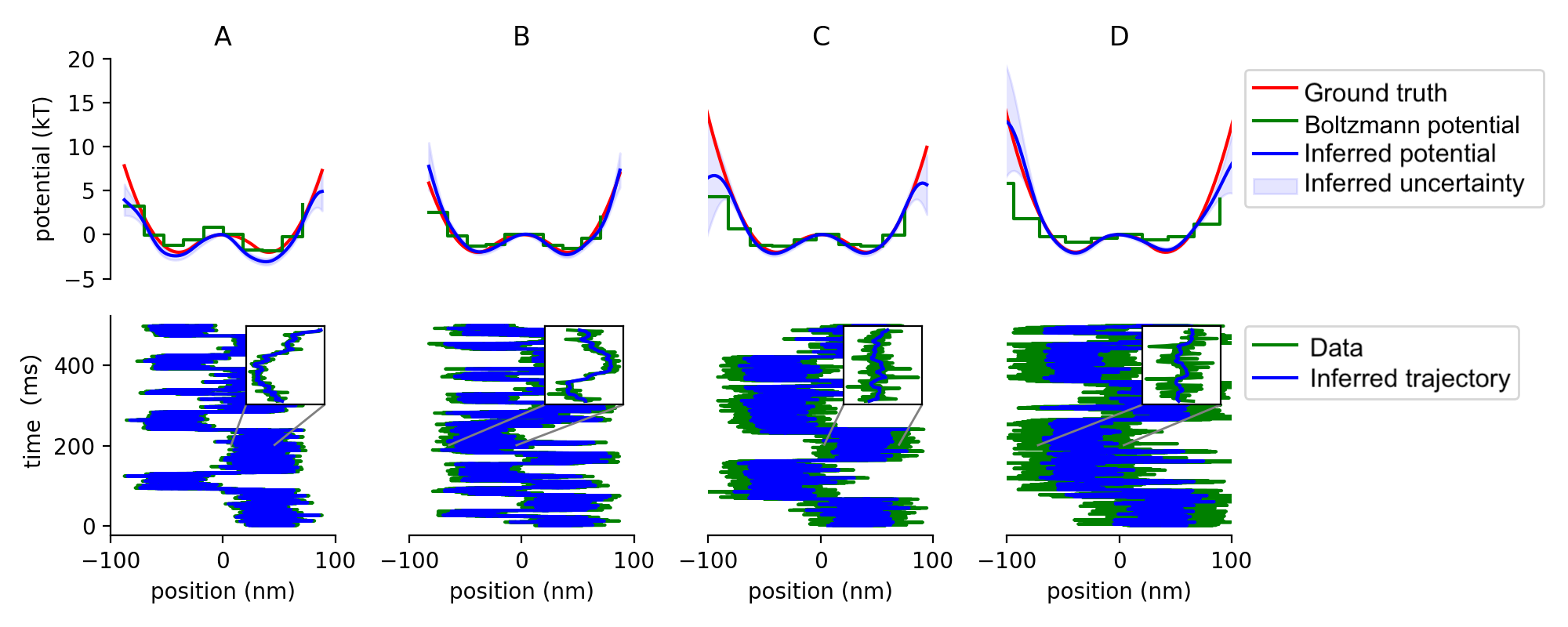}
    \caption{Demonstration on data from a double-well potential.  We analyze four data sets with increasing measurement noise. For each data set, we plot the inferred potential in the top row alongside  the ground truth and results of the Boltzmann method. We plot the inferred trajectory against the ground truth in the bottom row. For clarity, we zoom into a region of the trajectory (200~ms to 201~ms). Measurement noise is added by increasing the optical density of the ND filter. The optical densities of the sub-figures A, B, C, D are 0, 0.3, 0.5, 0.7 respectively.
    Each trace contains 50,000 data points.}
    \label{fig:double}
\end{figure*}

\begin{figure*} 
    \centering
    \includegraphics[width=\textwidth]{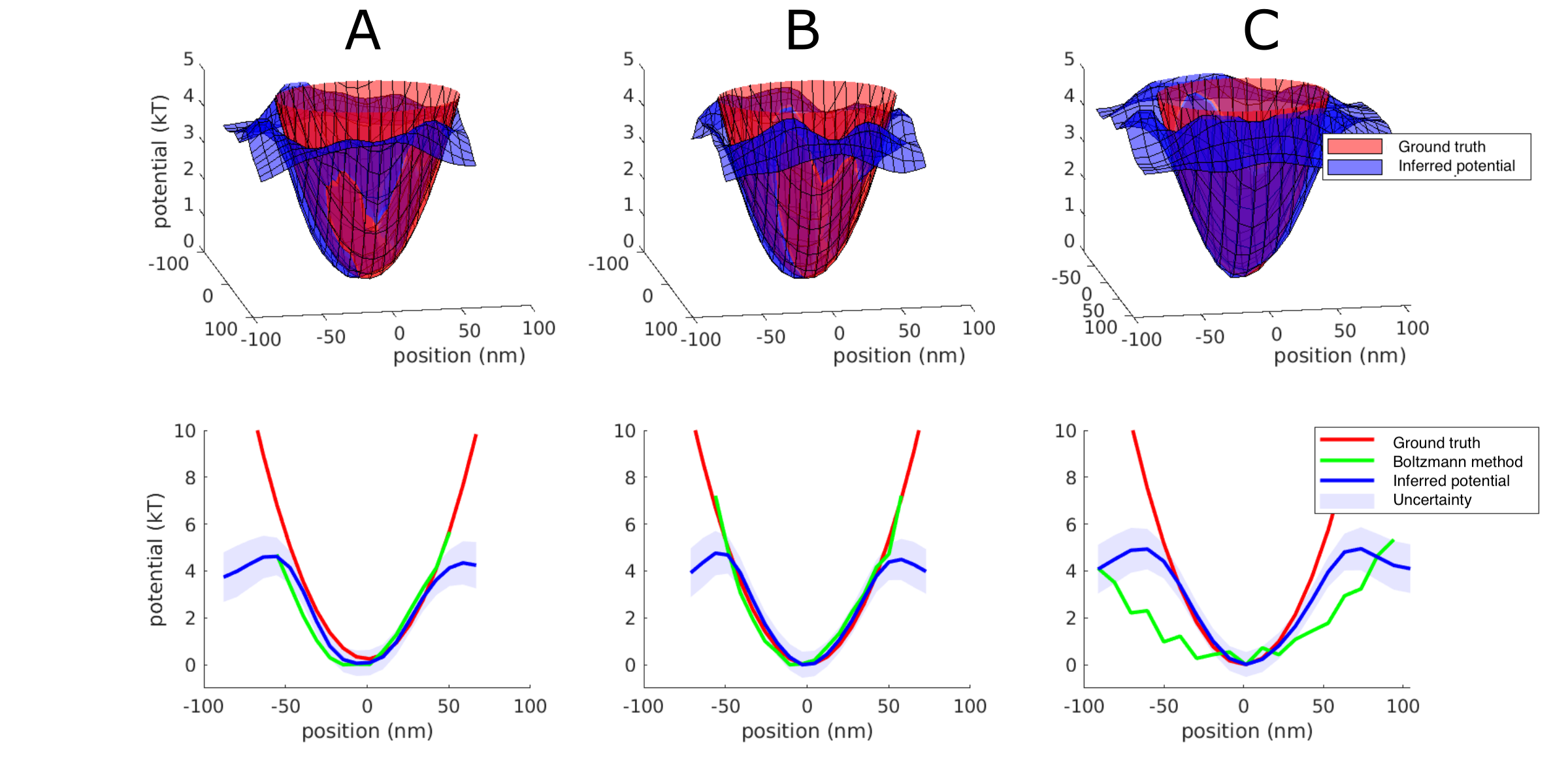}
    \caption{Demonstration on data from a 2D harmonic potential. Top row: three data sets with increasing measurement noise. Each column shows the inferred potential results along with the ground truth potential for a different data set. At the top we show the inferred potential and ground truth plotted in 2D. Bottom row: 1D slice taken through the middle of the potential. Measurement noise is added by increasing the optical density of the ND filter. The optical densities of ND filter used in the sub-figures A, B, C are 0.0, 0.3, 0.7 respectively.
    }
    \label{fig:2Dsingle}
\end{figure*}

\begin{figure*}
    \centering
    \includegraphics[width=\textwidth]{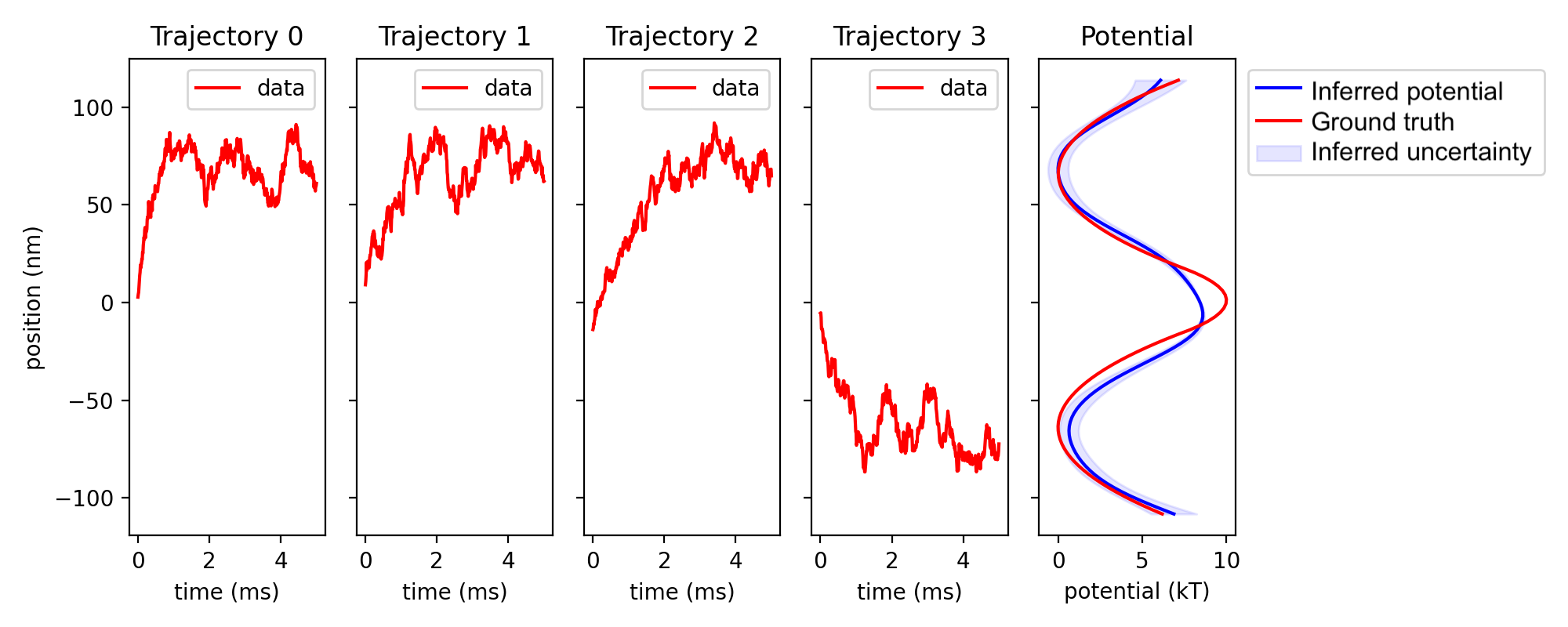}
    \caption{Demonstration of data from non-equilibrium trajectories. We reconstruct a potential by analyzing many short (500 data points) non-equilibrium trajectories. The left four panels show four of the 100 small data segments used to reconstruct the potential. Each trajectory starts at the top of the potential and rolls off to either side. The far right shows the inferred potential plotted with uncertainty overlaid on the ground truth potential and the potential inferred using the Boltzmann method. For comparison, the inferred and ground truth energy landscapes were shifted so that the lowest point is set to 0 $kT$.}
    \label{fig:noneq}
\end{figure*}

Determining potentials governing particle dynamics is of fundamental relevance to biology ~\cite{makarov2015shapes,wang2016nonlinear,wang1997configurational,wang2003energy,chu2013quantifying}, materials sciences~\cite{deringer2019machine,handle2018potential}, and beyond~\cite{perez2018forma, dudko2006intrinsic,preisler2004modeling, la2002potential}. 
For example, potentials provide reduced dimensional descriptions of dynamics along a reaction coordinate~\cite{wang2016nonlinear,wang2003energy,chu2013quantifying} and yield key estimates of thermodynamic and kinetic quantities~\cite{hanggi1990reaction, berezhkovskii2017mean, bessarab2013potential}. 
Shapes of energy landscapes also provide key insight into molecular function such as the periodic three-well potential of the $F_\textrm{o}F_{1}$-ATP synthase rotary motor ~\cite{wang1998energy, toyabe2012recovery} and the asymmetric, linearly periodic potentials responsible for kinesin's processivity~\cite{kolomeisky2007molecular}. 

In a different class of applications, fundamental experimental tests of statistical physics~\cite{proesmans2020finite,wu2009trajectory} often employ potentials with deliberately complex shapes created from feedback traps based on electrical~\cite{cohen2005control,gavrilov2013particle}, optical~\cite{kumar2018nanoscale,albay2018optical}, or thermal forces~\cite{braun2015single}, or  optically generated with phase masks~\cite{hayashi2008particle} or spatial light modulators~\cite{chupeau2020optimizing}.  

Inferring naturally occurring energy landscapes or verifying artificially created potentials demands a method free of {\it a priori} assumptions on the potential's shape.  This requirement rules out many commonly used methods devised for harmonic systems~\cite{neuman2004optical,berg2004power,jones2015book,gieseler2021optical} or alternative, otherwise-limited, methods to deduce potentials from data~\cite{reif2009fundamentals, turkcan2012bayesian, perez2018forma, wang2019machine, frishman2020learning, yang2021inference, stilgoe2021enhanced}. For example, some methods~\cite{reif2009fundamentals, turkcan2012bayesian} necessarily rely on binned data, relating potential energies to Boltzmann weights or average apparent force, thereby limiting the frequency of data in each bin and requiring that equilibrium be reached before data acquisition. Other methods assume stitched locally harmonic forms ~\cite{perez2018forma}.  Still others use neural networks~\cite{wang2019machine} to deduce potentials; the uncertainty originating  from measurement error and data sparsity is then not easily propagated to local uncertainty estimates over the inferred potential.

In previous work~\cite{bryan2020inferring}, we introduced a method starting from noiseless one-dimensional time series data to infer effective potential landscapes without binning, or assuming a potential form, or assuming equilibrium conditions, while admitting full posterior inference (and thus error bars or, equivalently, credible intervals) over any candidate potentials arising from sparse data.

Our method was, however, fundamentally limited to one dimension (because of the poor scaling of the computation with respect to the data set size). It also ignored measurement error and, thus, implications for uncertainty about the potential.

Here we introduce a method to infer potentials from noisy, multidimensional, non-equilibrium  data. We take advantage of tools from Bayesian nonparametrics to place priors over arbitrary shaped potentials. To do so, we introduce structured kernel interpolation~\cite{wilson2015kernel} to the Natural Sciences in order to circumvent the otherwise-prohibitive computational scaling of widely used Gaussian processes. As a result, our method can infer potential shapes from trajectories while meeting all the following criteria simultaneously: 1) no reliance on binning or pre-processing; 2) no assumed analytic potential form; 3)  inferences drawn from posteriors, allowing for spatially nonuniform uncertainties to be informed by local density of available data in specific regions of the potential (e.g., fewer data points around barriers); 4) treatment of multidimensional trajectories; 5) rigorous incorporation of measurement noise through likelihoods; and 6) compatible with non-equilibrium trajectories. No other existing method meets all six criteria simultaneously.

\textit{Methods.--}Our goal is to use noisy positional measurements, $\mtx{y}_{1:N}$, to infer all unknowns: 1) the potential at each point in space, $U(\cdot)$ (with $U(x)$ denoting the potential evaluated at $x$); 2) the friction coefficient, $\zeta$; 3) the magnitude of the measurement noise, $\sigma^2$ (under a Gaussian noise model); and 4) the actual position at each time, $\mtx{x}_{1:N}$. Toward achieving our goal, we construct a joint posterior probability distribution over all unknowns. As our posterior does not admit an analytic form, we devise an efficient Monte Carlo strategy to sample from it.

\textit{Data acquisition.--}We performed experiments using a feedback optical tweezer, whose details are given in the SI
and have been described in previous work~\cite{kumar2018nanoscale}. Briefly, we trap a silica bead of 1.5 \textmu m diameter using an optical tweezer, which creates a harmonic well without feedback. By applying feedback, we change the shape of the potential to a double well along one of the axes.
We use two different quadrant photodiodes (QPD) to measure the position ($\mtx{x}_{1},\mtx{x}_{2}$) of a bead at two different values of signal-to-noise ratio (SNR) simultaneously as explained in the supplemental information (SI)~\cite{supplementary}.
One detector has high SNR and is used for feedback to create the desired virtual potential~\cite{jun2012pre}; the other has an adjustable SNR and is used to explore inferences from measured signals with lower SNR.  We reduce the SNR in the other detector by placing neutral density (ND) filters of increasing optical density (OD) in front of it. Thus, we can use our method on the same trajectory over two different experimental SNRs and compare performance. We estimate the measurement noise and SNR in each detector from the noise floor of the power spectrum~\cite{supplementary}.

\textit{Dynamics.--}We describe the dynamics of the particle with an overdamped Langevin equation~\cite{zwanzig2001nonequilibrium},
\begin{subequations}
    \begin{align}
        \zeta \dot{\mtx{x}} &~=~ \mtx{f}(\mtx{x}) + \mtx{r}(t)\label{langevin}\\
        \mtx{f}(\mtx{x}) &~=~ -\mtx{\nabla} U(\mtx{x}),
    \end{align}\label{lang}
\end{subequations}
where $\mtx{x}(t)$ is the possibly multidimensional position coordinate at time $t$; $\dot{\mtx{x}}(t)$ is the velocity; $\mtx{f}(\mtx{x})$ is the force at position $\mtx{x}(t)$; and $\zeta$ is the friction coefficient.  The forces acting on the particle include positional forces $\mtx{f}(\mtx{x})$ expressed as the gradient of a conservative potential, $\mtx{f}(\mtx{x}) = -\mtx{\nabla}U(\mtx{x})$. The stochastic (thermal) force, $\mtx{r}(t)$, is defined as follows:
\begin{subequations}
    \begin{align}
        \langle \mtx{r}(t)\rangle&~=~\mtx{0} \\
        \langle r_i(t)r_j(t')\rangle&~=~2\zeta kT\delta(t-t')\delta_{ij}
    \end{align}
\end{subequations}
where $\langle\cdot\rangle$ denotes an ensemble average over realizations, $T$ is the temperature of the bath and $k$ is Boltzmann's constant. Under a forward Euler scheme~\cite{leveque2007finite} for Eq.~\eqref{langevin} with time points given by $t_n=n\dt$, each position, given its past realization, is sampled from a normal distribution 
\begin{align}
    \mtx{x}_{n+1}|\mtx{x}_n,\mtx{f}(\cdot),\zeta ~\sim~& \normal{\mtx{x}_n + \frac{\dt}{\zeta}\mtx{f}(\mtx{x}_n),~ \gamma^2\mtx{I}}. \label{dynamics}
\end{align}
In words, ``the position $\mtx{x}_{n+1}$ given quantities $\mtx{x}_n,\mtx{f}(\cdot),$ and $\zeta$ is sampled from a Normal distribution with mean $\mtx{x}_n + \frac{\dt}{\zeta}\mtx{f}(\mtx{x}_n)$ and variance $\gamma^2=\frac{2 \dt kT}{\zeta}\mtx{I}$."

As is typical for experimental setups, we use a Gaussian noise model and write
\begin{align}
    \mtx{y}_n |\mtx{x}_n,\sigma^2 ~\sim~& \normal{\mtx{x}_n,~ \sigma^2\mtx{I}} \label{noise}.
\end{align}
In words, the above reads ``$\mtx{y}_n$ given quantities $\mtx{x}_n,\sigma^2$ is drawn from a normal." Here $\sigma^2$ is the  measurement noise variance. In Eq.~\eqref{noise}, the measurement process is instantaneous, i.e., assumed to be faster than the dynamical time scales. Our choice of Gaussian measurement model here can be modified at minimal computational cost (e.g.,~\cite{hirsch2013stochastic}) if warranted by the data.

\textit{Probabilities.--}Next, from the product of the likelihood ($\prob{\mtx{y}_{1:N}| U(\cdot), \zeta, \mtx{x}_{1:N}, \sigma^2}$) and the prior ($\prob{U(\cdot), \zeta, \mtx{x}_{1:N}, \sigma^2}$), we obtain the posterior over all unknowns
\begin{align}
    \prob{U(\cdot), \zeta, \mtx{x}_{1:N}, \sigma^2 | \mtx{y}_{1:N}} &~\propto~ \prob{\mtx{y}_{1:N}| U(\cdot), \zeta, \mtx{x}_{1:N}, \sigma^2} \nonumber\\
    &\times \prob{U(\cdot), \zeta, \mtx{x}_{1:N}, \sigma^2}. \label{bayestheorem}
\end{align}

The likelihood is derived from the noise model provided in Eq.~\eqref{noise}. By contrast, the prior is informed by the Langevin dynamics, as we see by decomposing it as follows:
\begin{align}
    &\prob{U(\cdot), \zeta, \mtx{x}_{1:N}, \sigma^2} ~=~ \prob{\mtx{x}_{2:N}|\mtx{x}_{1}, U(\cdot),\zeta} \nonumber\\ &\times
    \prob{\mtx{x}_{1}|U(\cdot),\zeta}
    \prob{U(\cdot)}\prob{\zeta}\prob{\sigma^2}. \label{prior}
\end{align}
The first term on the right-hand side of Eq.~\eqref{prior} follows from Eq.~\eqref{lang}, while we are free to choose the remaining priors, $\prob{U(\cdot)}$, $\prob{\mtx{x}_{1}|U(\cdot),\zeta}$, $\prob{\zeta}$, and $\prob{\sigma^2}$.

Important considerations dictate the prior on the potential. First, the potential may assume any shape (and, as such, is modeled nonparametrically) although it should be smooth (i.e., spatially correlated). A Gaussian process (GP) prior~\cite{williams2006gaussian} allows us to sample continuous curves with covariance provided by a pre-specified kernel. However, naive GP prior implementations are computationally prohibitive, with time and memory requirements scaling as the number of data points cubed~\cite{bryan2020inferring,williams2006gaussian}. 

These size-scaling issues can be resolved by adopting a structured-kernel-interpolation GP (SKI-GP)~\cite{wilson2015kernel, wilson2015thoughts, titsias2009variational, gal2014variational} prior for the potential, $U(\cdot)$. The SKI-GP prior is a hierarchical structure, where the potential at all points is interpolated according to $M$ chosen inducing points at fixed locations, $\mtx{x}^u_{1:M}$, and where the values of the potential at the inducing points are themselves drawn from a GP, as detailed in the SI~\cite{supplementary}.
We note that under this model, we shift the focus from infering $U(\cdot)$ to infering $\mtx{u}_{1:M}$ from which we recover $U(\cdot)$ and $\mtx{f}(\cdot)$ with a modified kernel matrix~\cite{supplementary}.

Choices for priors on $\mtx{x}_1$, $\zeta$ and $\sigma^2$ are less critical and chosen for computational convenience alone~\cite{supplementary}.

\textit{Inference.--}As our posterior does not assume an analytical form, we devise an overall Gibbs sampling scheme~\cite{bishop2006pattern} to draw samples from it. Within this scheme, we start with an initial set of values for the parameters ($\zeta^{(0)}, \sigma^{(0)}, \mtx{x}_{1:N}^{(0)}, U(\cdot)^{(0)}$) and then iteratively sample each variable holding all others  fixed~\cite{geman1984stochastic} (see SI~\cite{supplementary}). 

\textit{Results.--}
We benchmark our method on experimental data on a double-well potential and show that we can accurately infer the shape of the potential. We then show that SKI-GP allows us to explore 2D time series data (previously infeasible due to large amounts of data using a naive GP). We finally apply our method to trajectories in a high-barrier landscape where traces are too short to reach equilibrium. A demonstration on data from a simple harmonic well and robustness tests over parameters of interest using simulated data can be found in the SI~\cite{supplementary}.

For testing the accuracy and effectiveness of GPs method, we simultaneously collected two measurements of each trajectory, one using a detector with low measurement noise and one using a second detector with higher measurement noise. We refer to the low-noise trajectory as the ``ground-truth'' trajectory, although it itself is subject to a small amount of measurement noise. For each experiment, we impose a potential on the particle using our feedback trap. We refer to this applied potential as the ``ground-truth" potential, although it may differ from the actual potential the particle experiences due to errors in the feedback trap setup, as well as experimental limitations such as drift. We use our ground-truth estimates to validate the accuracy of estimates made using our Gaussian processes on noisy and much shorter time-series data in the SI~\cite{supplementary}.

\textit{Double well.--}We analyzed data from a particle in a double-well potential. Results are shown in Fig.~\ref{fig:double}. Each column shows the inferred potential (top row) and inferred trajectory (bottom row) for each data set analyzed. We provide uncertainties and ground truth estimates for both the potential and trajectory. Additionally, for sake of comparison, we also show the potential estimated using the Boltzmann method~\cite{bryan2020inferring, reif2009fundamentals}. We highlight that the Boltzmann method does not provide trajectory estimates. By contrast, our method infers those positions obscured by noise. Fig.~\ref{fig:double} shows that the ground truth potential and trajectory fall within the estimated range even when the measurement noise is so large that the particle is occasionally seen in the wrong well (Fig.~\ref{fig:double}, top right panel). Both our method and the Boltzmann method slightly  overestimate the potential of the left well at the lowest noise level, because the (short) trajectory spends too much time in the right well, leaving the left well undersampled. 

\textit{2D single well.--}Next, we analyzed data from a particle in a 2D harmonic potential. Results are shown in Fig.~\ref{fig:2Dsingle}. For clarity, we do not show uncertainties, trajectories, or Boltzmann-method estimates for the 2D plot, but we do show them for a 1D potential slice. Despite the added complexity in inferring the potential in full 2D at once, our estimates fall within uncertainty in regions where data is appreciably sampled, even at high measurement noise.

\textit{Non-equilibrium trajectories--}One advantage of our method is that it does not rely on equilibrium assumptions. As such, we can analyze trajectories initiating from non-equilibrium conditions. To demonstrate this, we created data sets where the particle starts at the top of the potential well and ``rolls off'' to either side. The trajectories are short (5~ms), so that the particle does not reach equilibrium during the time trace. By including the likelihoods from 100 such trajectories into our posterior, we gain information on either side of the well and can recreate the full potential, even though each individual trajectory is initiated from the top of the potential.

In the first four panels of Fig.~\ref{fig:noneq}, we illustrate 4 of 100 trajectories used to reconstruct the potential. We note that all trajectories start from the top of the barrier (defined as $x=0$), and none of the trajectories fully sample both wells. Despite this extreme undersampling, our method is able to infer the height of the barrier to within 15\% accuracy (our method predicts an 8.6~$kT$ barrier; the barrier of the design potential is 10~$kT$). Our error bar at the top of the barrier in Fig.~\ref{fig:noneq} is artificially low because every trajectory initiates from the top. 

\textit{Discussion.--}Inferring potential landscapes is a key step toward providing a reduced dimensional description of complex systems~\cite{wang2019machine,chmiela2017machine,espanol2011obtaining,izvekov2005multiscale,manzhos2015neural}. Here, we go beyond existing methods by providing a means of obtaining potentials, amongst multiple other quantities, from time series data corrupted by measurement noise. We do so by efficiently learning the potential from the rawest form of data, point by point. That is, we achieve this without data pre-processing (e.g., binning), assuming an analytical potential form, nor requiring equilibrium conditions. 

As our method is Bayesian, it allows for direct error propagation to the final estimate of the inferred potential shape. In other words, our method differs from others assuming analytic potential forms~\cite{perez2021forma} or projection onto basis functions~\cite{frishman2020learning}, as well as methods relying on neural nets ~\cite{manzhos2015neural} that cannot currently propagate experimental uncertainty or provide error bars reflecting the amount of data informing the potential at a particular location. Importantly, unlike the Boltzmann method~\cite{reif2009fundamentals}, our method does not invoke any equilibrium assumption and can consider trajectories initiated from positions not sampled from an equilibrium distribution. This feature is especially relevant in studying landscapes with rarely sampled regions of space and, in particular, far from non-equilibrium.

As the method is general and the measurement noise model can be tuned, we can apply our method to mapping landscapes from force spectroscopy ~\cite{gupta2011experimental} or even single molecule fluorescence energy transfer~\cite{kilic2021extraction}, with applications to inferring protein conformational dynamics or binding kinetics~\cite{schuler2008protein,chung2018protein,sturzenegger2018transition}. In infering smooth potentials, we would move beyond the need to require discrete states inherent to traditional analyses paradigms such as hidden Markov models~\cite{rabiner1986introduction,sgouralis2017icon}.

\bibliographystyle{jabbrv_apsrev4-1}
\bibliography{main}

\end{document}


\preprint{AIP/123-QED}

\title[Supplementary Information]{Supplementary Information}

\author{J. Shepard Bryan IV*}
\affiliation{Department of Physics, Arizona State University, USA}

\author{Prithviraj Basak*}
\affiliation{Department of Physics, Simon Fraser University, CA}

\author{John Bechhoefer}
\email{johnb@sfu.ca}
\affiliation{Department of Physics, Simon Fraser University, CA}

\author{Steve Press{\'e}}
\email{spresse@asu.edu}
\affiliation{Department of Physics, Arizona State University, USA}
\affiliation{School of Molecular Sciences, Arizona State University, USA}

\maketitle

\section{Experimental apparatus}\label{Expt. apparatus}

\begin{figure}
	\centering
	\includegraphics[width=0.55\linewidth]{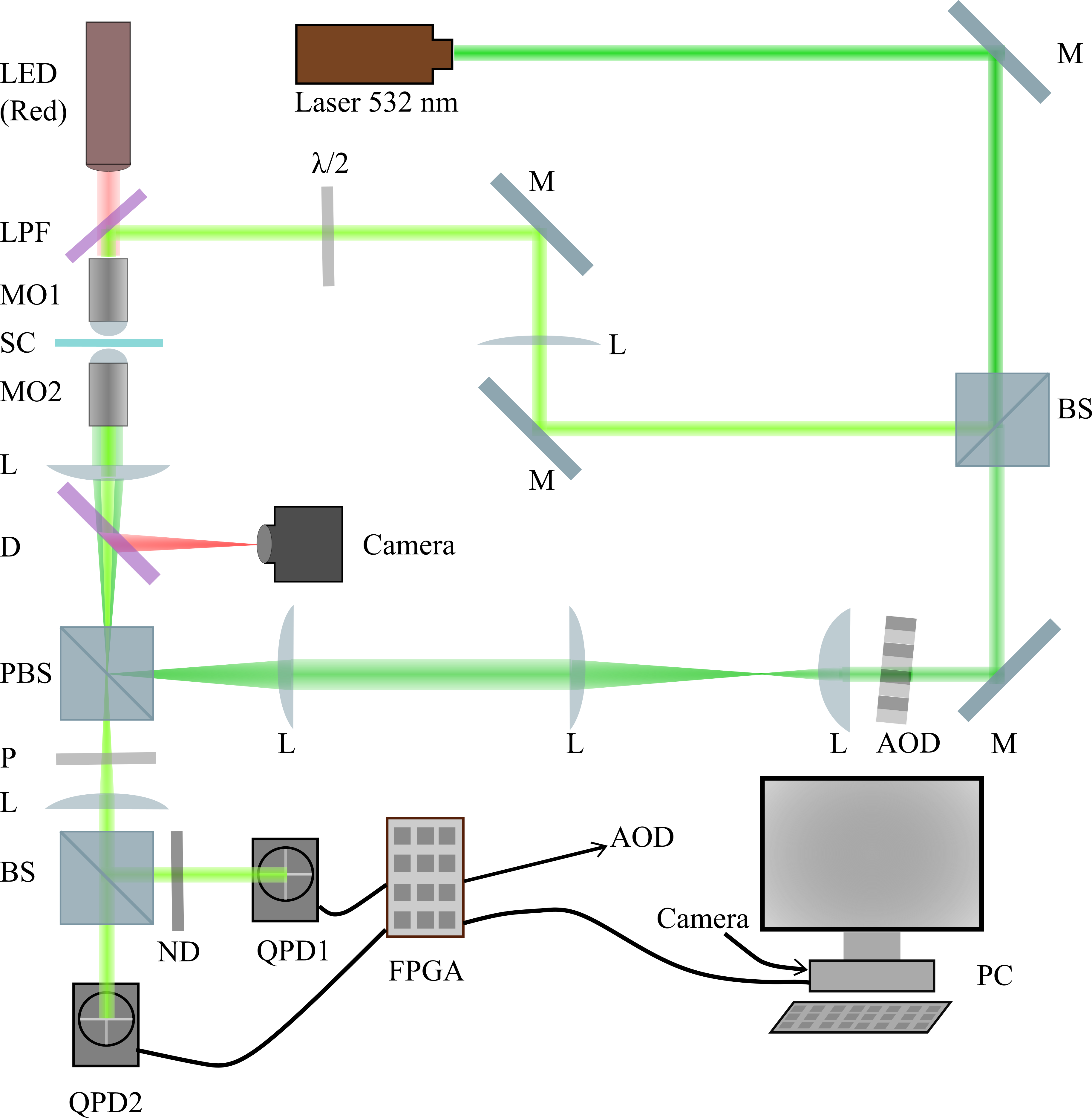}
	\caption{Schematic diagram of the experimental apparatus. M = mirror, L = lens, BS = beam splitter (non-polarizing), PBS = polarizing beam splitter, AOD = acousto-optic deflector, LPF = long-pass filter, MO = microscope objective, SC = sample chamber, SPF = short-pass filter, ND = neutral density filter, QPD = quadrant photodiode, \textlambda/2 = half-wave plate, P = linear polarizer.}
	\label{fig:schematics}
\end{figure}

The experiment is done on a modified version of the feedback optical tweezer used in various stochastic thermodynamics experiments involving virtual potentials~\cite{kumar2018nanoscale, kumar2020exponentially}. The schematics of the setup are provided in Fig.~\ref{fig:schematics}. We used a continuous-wave diode-pumped laser (H\"UBNER Photonics, Cobolt Samba, 1.5 W, 532 nm) and a custom-built microscope to construct the optical trap on a vibration-isolation table (Melles Griot). We split the laser into a trapping-beam and a detection-beam using a 90:10 beam splitter. The trapping beam then passes through a pair of acousto-optic deflectors (DTSXY-250-532, AA Opto Electronic), which allows us to deflect the beam in the orthogonal XY plane. We can steer the angle of the beam and also its intensity using analog voltage-controlled oscillators (DFRA10Y-B-0-60.90, AA Opto Electronic). We increase the beam-diameter using a two-lens system in telescopic configuration to overfill the back aperture of the trapping microscope objective (MO1 in Fig\ref{fig:schematics}). Then a 4f relay system images the steering point of the AOD on the back aperture of MO2. A water-immersion, high-numerical-aperture objective (MO2, Olympus 60X, UPlanSApo, NA = 1.2) is used for trapping 1.5 \textmu m diameter spherical silica beads (Bangs Laboratories) in aqueous solution (SC).

The detection beam is passed through a half-wave plate (\textlambda/2), so that its polarization is orthogonal to the trapping beam, to avoid unwanted interference. It is focused using a low-numerical-aperture microscope objective (MO1, 40X, NA = 0.4) antiparallel to the trapping objective MO2. The loosely focused detection beam (compared to the trapping beam) has a larger focal spot and offers a high linear range for the position detection. We can also adjust the detection plane using a 4f relay lens system. The forward scattered detection beam from the trapped bead is collected by the trapping objective MO2 and transmitted through the polarizing beam splitter (PBS). The PBS reflects the trapping beam and thus separates the detection beam. We also place another linear polarizer (P) after the PBS to minimize the amount of leaked trapping laser. The detection beam is then separated using a beam splitter on a pair of quadrant photodiodes (QPD, First Sensor, QP50-6-18u-SD2). We control the intensity of the detection beam on QPD1 by placing a neutral density filter (ND) of desired strength and thus control the signal-to-noise ratio (SNR) on QPD1.

A red LED (660nm, Thorlabs, M660L4) illuminates the trapped particle for imaging onto a camera. The light from the LED enters the detection object through a long-pass filter (LPF, cutoff wavelength 585~nm, Edmond Optics) which reflects the detection beam. The collected illumination light is separated by a short-pass filter (SPF, cutoff wavelength 600~nm, Edmond Optics) and reflected onto a camera (FLIR BFS-U3-04S2M-CS).

Using a LabView program, we digitize the analog signal from the pair of QPDs by a field programmable gate array (FPGA, National Instruments, NI PCIe-7857). The voltage signals are then calibrated as position signals using a QPD-AOD-camera method~\cite{kumar2018optical}. The FPGA runs the control loop and uses a programmed feedback rule to send the appropriate control signal to the AODs.



\section{Measurement noise}\label{Expt.noise}
\begin{figure}
	\centering
	\includegraphics[width=.8\linewidth]{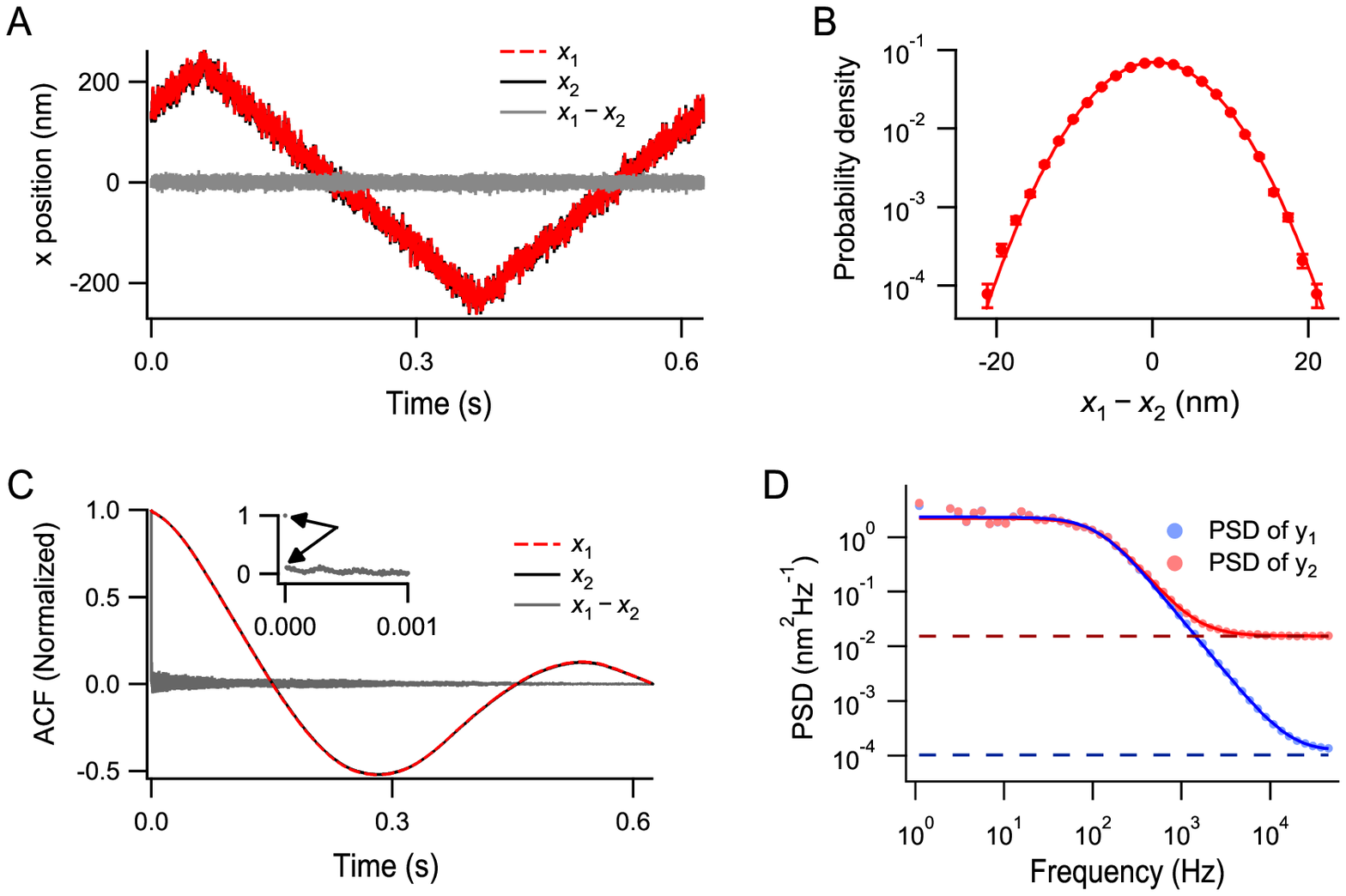}
	\caption{Measurement-noise calibration. A. The response of a bead's triangle wave motion is recorded as $\mtx{x}_{1}$ and $\mtx{x}_{2}$ by the two detectors. If the two detectors are accurately calibrated, the difference between the two signals $\mtx{x}_{1}- \mtx{x}_{2}$ in effect sums their noise. B. Histogram of the difference in the signal between the two detectors. The solid line is a fit of the data to a normal distribution, with mean $0.28 \pm 0.02$~nm and standard deviation $5.69\pm 0.02$~nm. C. Normalized autocorrelation function of the two signal and their difference.  As seen in the inset, the autocorrelation of $\mtx{x}_{1}- \mtx{x}_{2}$ decays to the residual correlation in one time step. D. Estimation of noise in each of $\mtx{y}_{1}$, $\mtx{y}_{2}$ trajectory by fitting the corresponding power spectral density. $\mtx{y}_{1}$ and $\mtx{y}_{2}$ are the signals of a bead trapped in harmonic potential recorded using the two detectors at different signal to noise ratio (SNR). The solid lines are the fits to the PSD using the aliased Lorentzian formula with an added noise term. The dashed lines are the estimated noise in each of the PSD. We find the variance of the noise by integrating the noise term over the entire frequency region and estimate their signal-to-noise ratios to be 95.5 and 1.7, respectively.} 
	\label{fig:noise_hist}
\end{figure}
Here we discuss the estimation of measurement noise for our experimental setup. Briefly, we obtain an estimate of the noise by recording two simultaneous trajectories of a particle and finding the difference between the trajectories. As both measured trajectories should be equal in the limit of zero noise, we can interpret any deviation between the measured trajectories as coming from noise.

We start by applying a triangle wave voltage of frequency 1.6~Hz to the AOD, to move a bead linearly along the x-axis. We record the beads motion at 100 kHz using two simultaneous detectors giving output $\mtx{x}_{1}(t)$ and $\mtx{x}_{2}(t)$, respectively.
We demonstrate this idea in Fig.~\ref{fig:noise_hist}A. Figure~\ref{fig:noise_hist}B shows that the difference $\mtx{x}_{1} - \mtx{x}_{2}$ follows a Gaussian distribution. The mean of the Gaussian distribution shown in Fig.~\ref{fig:noise_hist}B, has a nonzero value of $0.28 \pm 0.02$~nm, which arises from nonlinear calibration errors in the two detectors. In Fig.~\ref{fig:noise_hist}C, we verify that the autocorrelation is flat, indicating that each position measurement has independent measurement noise.

To calculate the variance of the measurement noise, we denote the standard deviation of the detectors $\sigma_1$ and $\sigma_2$, respectively. For Gaussian distributed measurement noise $\mtx{x}_{1} - \mtx{x}_{2}$ is also Gaussian, with variance equal to $\sigma_{\mtx{x}_1}^2 + \sigma_{\mtx{x}_2}^2$.

We can fit the power spectral density of $\mtx{x}_1-\mtx{x}_2$ with the aliased Lorentzian expression for discretely sampled times series~\cite{berg2004power}. Since the measurement noise is Gaussian, we add a noise term to the fitting function. Thus we can estimate the noise of each trajectory by integrating the noise term over the frequency domain. From this, we estimate standard deviations of noise in $\mtx{x}_{1}$ and $\mtx{x}_{2}$ as $3.1 \pm 0.3$~nm and $3.8 \pm 0.4$~nm, respectively.

\section{Priors}\label{priors}

For $\prob{\mtx{x}_1|U(\cdot),\zeta}$, we are free to choose any prior and select
\begin{align}
    \prob{\mtx{x}_1|U(\cdot),\zeta} = \normal{\mtx{x}_1;0,\Theta^2}.\label{x0prior}
\end{align}
For the friction, $\zeta$, we select a gamma distribution for which support on the positive real axis is assured. That is,
\begin{align}
    \zeta ~\sim~& \Gammapdf{\alpha_\zeta, \beta_\zeta}\label{frictionprior}.
\end{align}
Lastly, for the noise variance, $\sigma^2$, we place an inverse-gamma prior
\begin{align}
    \sigma^2 ~\sim~& \invgamma{\alpha_{\sigma^2}, \beta_{\sigma^2}}\label{noiseprior}.
\end{align}
The inverse-gamma prior is chosen because it is conjugate to the likelihood, meaning that we may directly sample from the posterior constructed by the prior multiplied by the likelihood ~\cite{bishop2006pattern}. The variables, $\Theta$, $\alpha_\zeta$, $\beta_\zeta$, $\alpha_{\sigma^2}$, and $\beta_{\sigma^2}$ are hyperparameters that we are free to choose, and whose impact on the ultimate shape of the posterior reduces as more data are collected~\cite{gelman2013bayesian,sivia2006data}.

\section{Modified kernel matrix}\label{modified kernel matrix}

Here we show how to calculate the force at an arbitrary test location using the structured kernel interpolation Gaussian process. We first derive the force for one-dimensional potentials exactly and then briefly explain how to generalize to higher dimensions. In order to calculate the force given $\mtx{u}_{1:M}$, we write
\begin{subequations}
    \begin{align}
        f(x) ~=~& - \frac{d}{dx} U(x)\\
        ~=~& - \frac{d}{dx} \mtx{K^\dagger}\mtx{K}^{-1}\mtx{u}_{1:M}\\
        ~=~& \mtx{K^*}\mtx{K}^{-1}\mtx{u}_{1:M}\\
        \mtx{K^*} ~=~& - \frac{d}{dx} \mtx{K^\dagger}\label{inducing}\\
        K^*_{ij} ~=~& - \frac{d}{dx} K(x, x^u_m)\\
        ~=~& \frac{h^2}{\ell^2}\left(x-x^u_m\right)\exp\left(-\frac{1}{2}\left(\frac{x-x^u_m}{\ell}\right)^2\right) \,,
    \end{align}
\end{subequations}
where $\mtx{K^*}$ is now the covariance between the potential evaluated at the inducing points and the force evaluated at a test location, and we have assumed the kernel follows the familiar squared exponential form~\cite{williams2006gaussian}.

To generalize to higher dimensions, we will need to calculate the force in each dimension separately. For example, to find the force in the $k$ direction at a test point, $\mtx{x}$,
\begin{align}
    f^k(\mtx{x}) ~=~& -\frac{d}{dx^k}U(\mtx{x})\\
    ~=~& - \frac{d}{dx^k} \mtx{K^\dagger}\mtx{K}^{-1}\mtx{u}_{1:M}\\
    ~=~& \mtx{K^{*k}}\mtx{K}^{-1}\mtx{u}_{1:M}\\
    K^{*k}_{ij}~=~& \frac{h^2}{\ell^2}\left(x^k-x^{u,k}_m\right)\exp\left(-\frac{1}{2}\left|\frac{\mtx{x}-\mtx{x}^u_m}{\ell}\right|^2\right).
\end{align}

\section{Conditional probabilities for Gibbs sampling}\label{Conditional probabilities for Gibbs sampling}

Here, we show the conditional probabilities used for our Gibbs sampling algorithm.

\subsection{Positions}\label{sec positions}

The distribution for $\mtx{x}_{1:N}$ is simple. If we sample each $\mtx{x}_n$ one at a time,
\begin{align}
    \prob{\mtx{x}_1|\mtx{u}_{1:M},\mtx{x}_{2:N}} ~\propto~& \normal{\mtx{x}_1;~ 0,~ \frac{2\tau kT}{\zeta}}\\
    &~\times~ \normal{\mtx{x}_2;~ \mtx{x}_1+\frac{\tau}{\zeta}\mtx{f}(\mtx{x}_1),~ \frac{2\tau kT}{\zeta}}\notag\\
    &~\times~ \normal{\mtx{y}_1;~ \mtx{x}_1,~ \sigma^2}\notag\\
    \prob{\mtx{x}_n|\mtx{u}_{1:M},\mtx{x}_{1:n-1}} ~\propto~& \normal{\mtx{x}_n;~ \mtx{x}_{n-1} + \frac{\tau}{\zeta}\mtx{f}(\mtx{x}_{n-1}),~ \frac{2\tau kT}{\zeta}}\\
    &~\times~ \normal{\mtx{x}_{n+1};~ \mtx{x}_n+\frac{\tau}{\zeta}\mtx{f}(\mtx{x}_n),~ \frac{2\tau kT}{\zeta}}\notag\\
    &~\times~ \normal{\mtx{y}_n;~ \mtx{x}_n,~ \sigma^2}\notag\\
    \prob{\mtx{x}_N|\mtx{u}_{1:M},\mtx{x}_{1:N-1}} ~\propto~& \normal{\mtx{x}_{N};~ \mtx{x}_{N-1} + \frac{\tau}{\zeta}\mtx{f}(\mtx{x}_{N-1}),~ \frac{2\tau kT}{\zeta}}\\
    &~\times~ \normal{\mtx{y}_N;~ \mtx{x}_N,~ \sigma^2}\notag.
\end{align}
Although these equations look Gaussian, they are not (except the last one), since we have to remember that $\mtx{f}_n=\mtx{f}(\mtx{x}_n)$ is a function of $\mtx{x}_n$. Thus, direct sampling is not possible, and we can only sample positions using a Metropolis-Hastings algorithm~\cite{bishop2006pattern}.

\subsection{Potential}\label{potential}

Following the logic outlined in our previous manuscript~\cite{bryan2020inferring}, we can infer the potential. For now, we focus on one-dimensional data. 

We fully write the prior on positions, $\mtx{x}_{2:N}|U(\cdot),\mtx{x}_1,\zeta$ as
\begin{align}
    \prob{\mtx{x}_{2:N}|\mtx{f}(\cdot),\mtx{x}_1} ~=~& \normal{\tau\mtx{v}_{1:N-1};~ \frac{\tau}{\zeta}\mtx{f}_{1:N-1},~ \frac{2\tau kT}{\zeta}\mtx{I}}\\
    ~\propto~& \normal{\mtx{f}_{1:N-1};~ \zeta\mtx{v}_{1:N-1},~ \frac{2\zeta kT}{\tau}\mtx{I}}
\end{align}
where $\mtx{f}_n=\mtx{f}(\mtx{x}_n)$ and $v_{n}=(\mtx{x}_{n+1}-\mtx{x}_n)/\tau$. Substituting in Eq. \ref{inducing}, we get,
\begin{align}
    ~=~& \normal{\mtx{K^*}\mtx{K}^{-1}\mtx{u}_{1:M};~ \zeta\mtx{v}_{1:N-1},~ \frac{2\zeta kT}{\tau}\mtx{I}}\\
    ~\propto~& \exp\left( -\frac{1}{2} \left(\zeta\mtx{v}_{1:N-1} - \mtx{K^*}\mtx{K}^{-1}\mtx{u}_{1:M}\right)^T \frac{\tau}{2\zeta kT}\mtx{I} \left(\zeta\mtx{v}_{1:N-1} - \mtx{K^*}\mtx{K}^{-1}\mtx{u}_{1:M}\right)\right)\\
    ~\propto~& \exp\left( -\frac{1}{2} \frac{\tau}{2\zeta kT}\mtx{u}_{1:M}^T\mtx{K}^{-1}\mtx{K^*}{}^{T} \mtx{K^*}\mtx{K}^{-1}\mtx{u}_{1:M}
    + \frac{1}{2}\frac{2\tau}{2kT}\mtx{v}_{1:N-1}^T \mtx{K^*}\mtx{K}^{-1}\mtx{u}_{1:M} \right)\\
    ~\propto~& \exp\left( -\frac{1}{2}\left(\mtx{u}_{1:M}-\mtx{M}^{-1}\mtx{b}\right)^T\mtx{M}\left(\mtx{u}_{1:M}-\mtx{M}^{-1}\mtx{b}\right)\right)\\
    ~\propto~& \normal{\mtx{u}_{1:M};~ \mtx{M}^{-1}\mtx{b},~ \mtx{M}^{-1}}
\end{align}
where we have used matrix completing the square, with
\begin{align}
    \mtx{b} ~=~& \frac{\tau}{2 kT} \mtx{K}^{-1}\mtx{K^*}{}^{T}\mtx{v}_{1:N-1}\\
    \mtx{M} ~=~& \frac{\tau}{2\zeta kT}\mtx{K}^{-1}\mtx{K^*}{}^{T} \mtx{K^*}\mtx{K}^{-1}.
\end{align}

Combining with the prior, we get
\begin{align}
    \prob{u_{1:M}|\mtx{x}_{1:N}} ~\propto~& \normal{\mtx{u}_{1:M};~ \mtx{M}^{-1}\mtx{b},~ \mtx{M}^{-1}} \normal{\mtx{u}_{1:M};~ \mtx{0},~ \mtx{K}}\\
    ~\propto~& \normal{\mtx{u}_{1:M};~ \left(\mtx{K}^{-1}+\mtx{M}\right)^{-1}\left( \mtx{K}^{-1}\mtx{0} + \mtx{M}\mtx{M}^{-1}\mtx{b}\right),~ \mtx{K}^{-1}+\mtx{M}}\\
    ~=~& \normal{\mtx{u}_{1:M};~ \tilde{\mtx{\mu}},~ \tilde{\mtx{K}}}\label{gpfinal}\\
    \tilde{\mtx{\mu}} ~=~& \frac{\tau}{2kT}  \left(\mtx{K}^{-1} +\frac{\tau}{2\zeta kT}\mtx{K}^{-1}\mtx{K^*}{}^{T} \mtx{K^*}\mtx{K}^{-1}\right)^{-1} \mtx{K}^{-1}\mtx{K^*}{}^{T}\mtx{v}_{1:N-1}\\
    \tilde{\bm{K}} ~=~& \left(\mtx{K}^{-1} +\frac{\tau}{2\zeta kT}\mtx{K}^{-1}\mtx{K^*}{}^{T} \mtx{K^*}\mtx{K}^{-1}\right)^{-1}.
\end{align}

For multidimensional trajectories, we separate the likelihood into the forces experienced along each trajectory,
\begin{align}
    \prob{\mtx{x}_{2:N}|\mtx{f}(\cdot),\mtx{x}_1} ~\propto~& \prod_d^D \normal{\mtx{f}^d_{1:N-1};~ \zeta\mtx{v}^d_{1:N-1},~ \frac{2\zeta kT}{\tau}\mtx{I}},
\end{align}
where $d$ is the dimension index and $D$ is the number of dimensions. A convenient choice of indexing allows us to sample the potential directly, even with multidimensional trajectories. It consists of ``flattening out'' our multidimensional force, velocity and kernel arrays into one-dimensional vectors, and ``flattening out'' the multidimensional kernel matrices into a matrix.

The convenient choice is as follows: Let the first $N$ indices of all variables be reserved for values concerning the $N$ measurements in the first dimension, then the $N+1$ through $2N$ indices be reserved for values concerning the $N$ measurements in the second dimension, and continue this pattern until all $D\times N$ measurements have been accounted for. For example, define $\mtx{f}$ to be a vector such that $\mtx{f}_1$ is the force at the first time level along the first dimension, $\mtx{f}_2$ is the force at the first time level along the second dimension, ..., $\mtx{f}_{D+1}$ is the force at the first time level along the second dimension, $\mtx{f}_{D+2}$ is the force at the second time level along the second dimension, and so forth. By utilizing this pattern for $\mtx{f}$, $\mtx{v}$, and $\mtx{K}^*$, we can sample the potential directly using Eq.~\eqref{gpfinal}.

\subsection{Friction coefficient}\label{sec Friction coefficient}

The conditional probability for the friction coefficient is the product of Eqs.~\eqref{dynamics},~\eqref{x0prior}~and~\eqref{frictionprior}
\begin{align}
    \prob{\zeta|\mtx{x}_{1:N},U(\cdot)} ~\propto~& \Gammapdf{\zeta;\alpha_\zeta, \beta_\zeta}\normal{\mtx{x}_1;~ 0,~ \frac{2\tau kT}{\zeta}}\nonumber\\
    &~\times~ \prod_{n=2}^N  \normal{\mtx{x}_n;~ \mtx{x}_{n-1} + \frac{\tau}{\zeta}\mtx{f}(\mtx{x}_{n-1}),~ \frac{2\tau kT}{\zeta}} \,,
\end{align}
which cannot be simplified into any known elementary distribution. We therefore sample, $\zeta$ using a Metropolis Hastings algorithm~\cite{bishop2006pattern}.

\subsection{Measurement noise}\label{sec measurement noise}

As our inverse gamma prior for $\sigma^2$, Eq.~\eqref{noiseprior} is conjugate to our likelihood, our conditional probability for $\sigma^2$ can be simplified to an inverse gamma distribution~\cite{gelman2013bayesian,sivia2006data}
\begin{align}
    \prob{\sigma^2|\mtx{y}_{1:N},\mtx{x}_{1:N}} ~=~& \invgamma{\alpha_{\sigma^2} + \frac{N}{2},~ \beta_{\sigma^2} + \frac{1}{2}\sum|\mtx{x}_n-\mtx{y}_n|^2}.
\end{align}

\section{Boltzmann method}\label{Boltzmann method}

In the Results section, we will compare to the Boltzmann method~\cite{reif2009fundamentals}. The Boltzmann method, as opposed to the other existing methods, has the advantage that it is physically intuitive (it is derived from thermodynamics).  It does not assume a potential shape a priori and therefore can be used for non-harmonic potentials. Its main limitation, in not treating measurement noise, is a limitation of all other competing methods.

The Boltzmann method uses the Boltzmann distribution from thermodynamics, which relates the potential in a region of space to the fraction of time that the particle will be seen in that region,
\begin{align}
    U_i &~\propto~- kT \log(p_i) \label{boltzmann}
\end{align}
where $U_i$ is the potential of region $i$ and $p_i$ is the fraction of the time that the particle spent in region $i$.

A limitation of the Boltzmann method is that the presence of measurement noise implies that the fraction of time that a particle \textit{was seen} in a region, $p_i$, does not equal the fraction of time that the particle \textit{was in} the region. As a consequence, measurement noise will smear the shape of the inferred potential over the range of the measurement noise (see Fig.~\ref{fig:2Dsingle}).

\section{Demonstration on simple harmonic well}\label{Demonstration on simple harmonic well}
\begin{figure*}
    \centering
    \includegraphics[width=\textwidth]{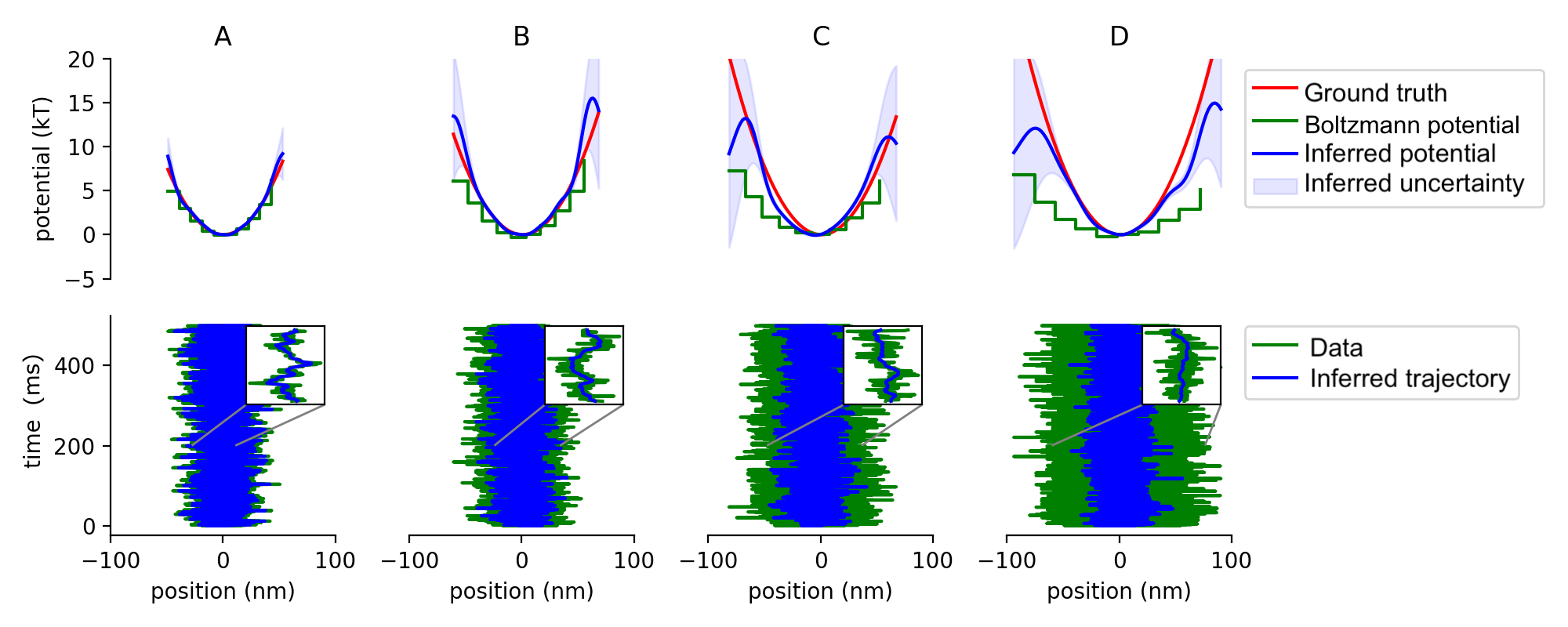}
    \caption{Demonstration on data from a harmonic potential.  Here we analyze four data sets with increasing measurement noise.
    For each data set, we plot the inferred potential in the top row along with the ground truth and the results of the Boltzmann method. We plot the inferred trajectory with uncertainty against the ground truth in the bottom row. For clarity, we zoom into a region of the trajectory (200~ms to 201~ms).
    Measurement noise is added by increasing the optical density of the ND filter. The optical densities of the sub-figures A, B, C, D are 0, 0.3, 0.5 and 0.6 respectively.
    Each trace contains 50,000 data points.}
    \label{fig:single}
\end{figure*}

\begin{figure*}
    \centering
    \includegraphics[width=\textwidth]{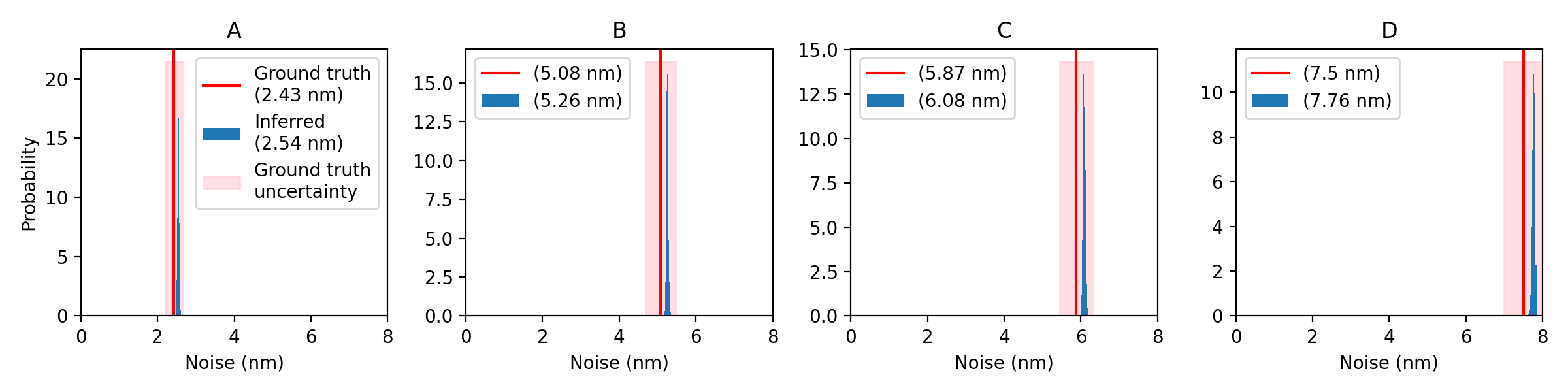}
    \caption{Inferred noise levels from experiments on harmonic well.  We analyze four data sets with increasing measurement noise. For each data set, we plot the probability density function of the inferred magnitude of the noise against a vertical line representing the best estimate inferred using the calibration techniques outlined earlier, with a shaded pink region representing the uncertainty in the calibrations establishing ``ground truth." Measurement noise is added by increasing the optical density of the ND filter, thereby decreasing the light intensity incident on the QPD. The optical densities of sub-figures A, B, C, D are 0, 0.3, 0.5 and 0.6, respectively.
    Each trace contains 50,000 data points.}
    \label{fig:single_noise}
\end{figure*}

We started by analyzing trajectories from a simple harmonic well. Results are shown in Fig.~\ref{fig:single}. Each column shows the inferred potential (top row) and inferred trajectory (bottom row) for each data set analyzed. We provide uncertainties and ground truth estimates for both the potential and trajectory. Additionally, for sake of comparison, we also show the potential estimated using the Boltzmann method~\cite{bryan2020inferring, reif2009fundamentals} (discussed earlier). We highlight that the Boltzmann method does not provide trajectory estimates. By contrast, our method infers those positions obscured by noise.

Figure~\ref{fig:single}  shows that the ground truth potential and trajectory fall within our error bars (credible interval) for all data sets up to ND=7. At ND=7, the inferred potential develops bumps where it is unable to infer the trajectory accurately, resulting in a 2~nm shift of the potential well minimum. On the other hand, the Boltzmann method (see earlier discussion) does well in the low-noise case (Fig.~\ref{fig:single}, left column) but fails as the noise introduced grows (Fig.~\ref{fig:single}, right column). This is expected, as measurement noise broadens the position histogram and thereby the potential. Using our method, the estimate of the potential drops at the  edges of the spatial region sampled by the particle, since the high-potential edges are rarely visited by the particle. We thus have insufficient information through the likelihood to inform those regions, and the  inferred potential reverts back to the prior (set at 0, as described earlier).

In addition to infering estimates for the potential energy landscape and trajectory, we also estimate the magnitude of the measurement noise for each experiment. Figure~\ref{fig:single_noise} shows the inferred measurement noise magnitudes obtained from our method for each single-well experiment, with the mean of our PDFs within 5\% of the ``ground truth" measurement, for each data set analyzed.

\section{Robustness tests on simulated data}\label{robustness test using simulated data}

In this section, we demonstrate our method on data simulated with a single-well potential analyzed under different circumstances to probe the robustness with respect to the number of data points and to measurement noise.

\subsection{Varying the number of data points}

\begin{figure}
    \centering
    \includegraphics[width=\textwidth]{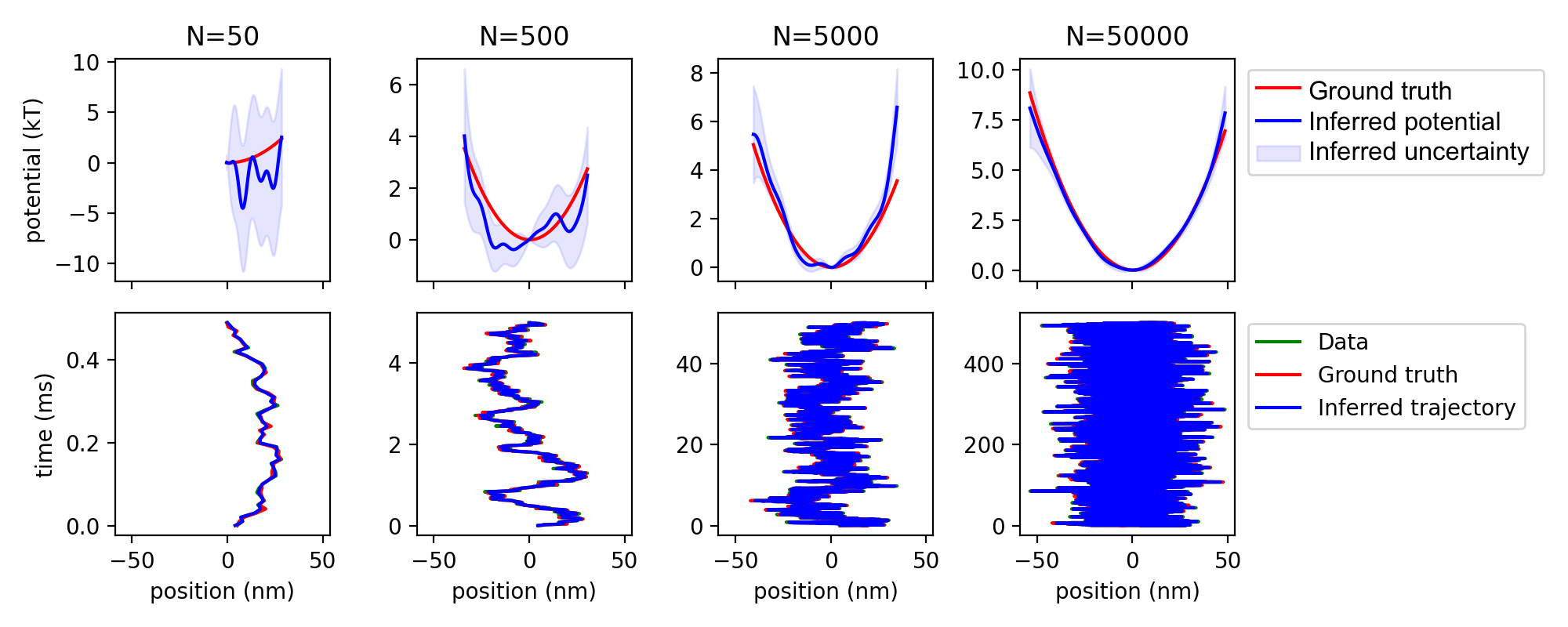}
    \caption{Robustness test with respect to number of data points. Here we analyze four data sets with a different number of data points. We show that increasing the number of data points increases our accuracy. Note that the y axis of each panel in the second row is different and corresponds to the length of the inferred trajectory.}
    \label{fig:vary_num_data}
\end{figure}

We tested our method's robustness with respect to the number of data points. Figure~\ref{fig:vary_num_data} shows the results. Trivially, when the trajectory is so short that the particle does not travel across the well ($N=50$ as in Fig.~\ref{fig:vary_num_data} left panel), we cannot infer the shape of the potential. When the particle samples the entire well ($N=500$ as in Fig.~\ref{fig:vary_num_data} second panel), we infer the general shape of the potential, but the inference is highly impacted by small stochastic anomalies.  See for example the far right of the potential, where a few higher- and lower-than-expected thermal kicks at the right side of the well caused our method to infer a second well. For reasonable inference, a few thousand points suffice ($N=5000$ in Fig.~\ref{fig:vary_num_data}, third panel). When there are tens of thousands of data points ($N=50,000$ in Fig.~\ref{fig:vary_num_data}), our inferred potential almost exactly overlaps with the ground truth.

\subsection{Varying the noise}

\begin{figure}
    \centering
    \includegraphics[width=\textwidth]{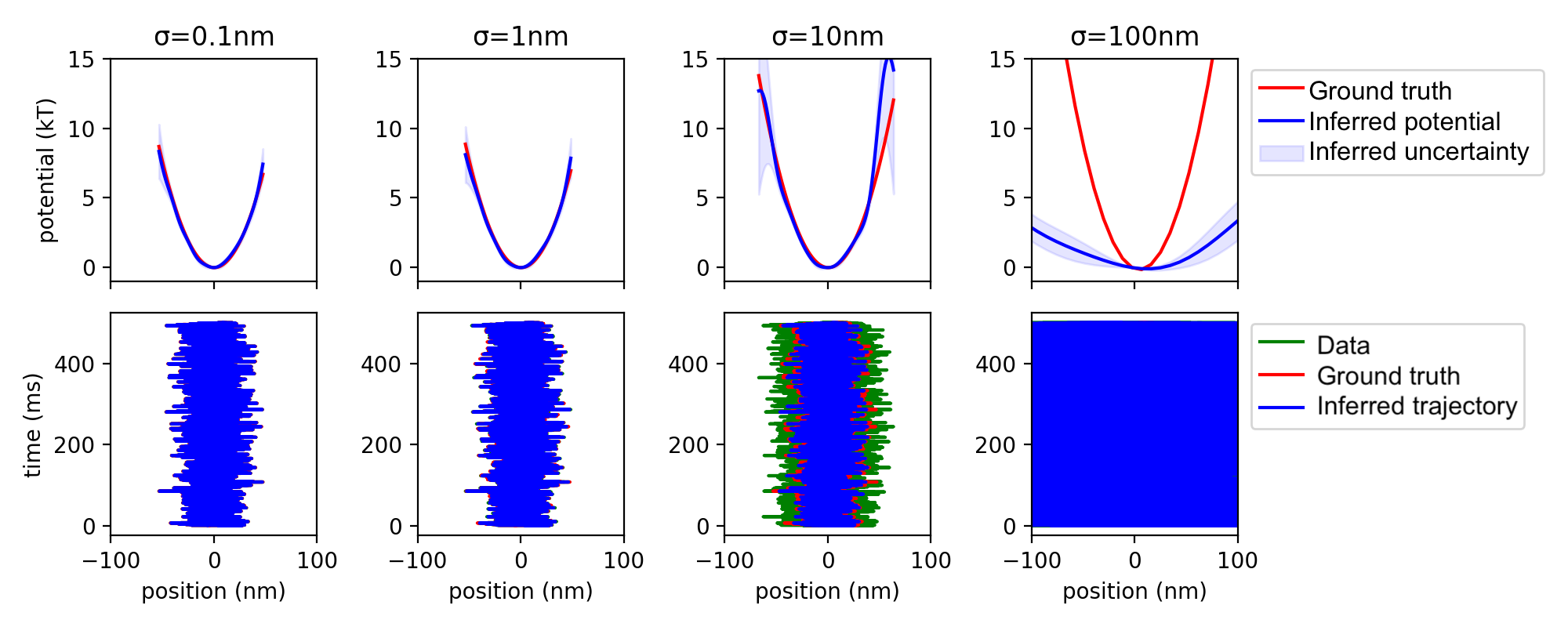}
    \caption{Demonstration on data from non-equilibrium trajectories. Here we analyze four data sets with different measurement noise magnitudes. We show that our method is robust with respect to measurement noise.}
    \label{fig:vary_noise}
\end{figure}

We tested our method's robustness with respect to measurement noise. Figure~\ref{fig:vary_noise} shows the results. Our method was robust to measurement noise for the four values of measurement noise variance chosen, but could not reproduce the potential energy landscape when the magnitude of the noise was of the same order as the maximum range of the particle. Even so, when the magnitude of the noise was up to 10\% the maximum range of the particle ($\sigma=10$ nm), we nonetheless inferred the potential energy landscape accurately.

\bibliographystyle{jabbrv_apsrev4-1}
\bibliography{bib}